# Probing Electro-Magnetic Field Enhancement in 3D Plasmonic Nanopores Using DNA-PAINT and Nanorulers

German Lanzavecchia[1,2], Anastasiia Sapunova[1,3], Alan Szalai[4], Shukun Weng[1,3], Ali Douaki[1,2], Makusu Tsutsui[5], Roman Krahne[1], Guillermo Acuña[4*], Denis Garoli[1,2*]

1. Optoelectronics, Istituto Italiano di Tecnologia, 16163 Genova, Italy;
2. Dipartimento di Scienze e Metodi dell'Ingegneria, Università degli Studi di Modena e Reggio Emilia, 43122 Reggio Emilia, Italy
3. Università degli Studi di Milano-Bicocca, 20126, Milano, Italy
4. Department of Physics, University of Fribourg, Fribourg CH-1700, Switzerland
5. SANKEN, The University of Osaka, 495 Osaka, Ibaraki 567-0047, Japan

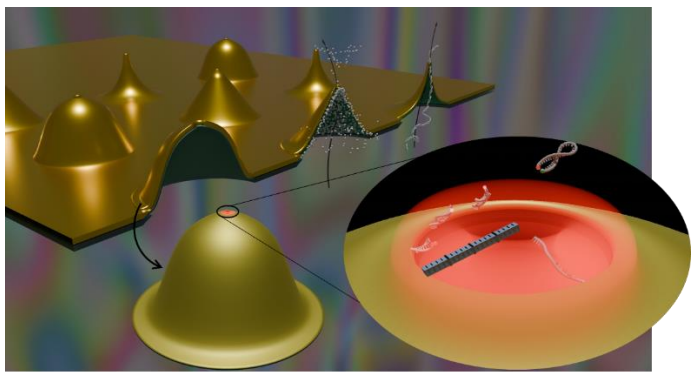

TOC figure - Schematic representation of DNA-based optical probing in plasmonic nanopores. Fluorescent DNA strands positioned at controlled distances from the metallic surface act as nanorulers, while DNA-PAINT events visualize local field enhancement at the nanopore tip. Together, these strategies reveal distance-dependent emission governed by the interplay of plasmonic enhancement and quenching.

## Abstract

Plasmonic nanopores combine nanofluidic confinement with electromagnetic field enhancement, enabling optical interrogation of single molecules in sub-wavelength volumes. Yet, direct optical readout within these metallic geometries has remained challenging due to fluorescence quenching near the surface. Here, we implement DNA-PAINT as a molecular reporter of local optical fields inside plasmonic nanopores. Transient hybridization of fluorescent imager strands at the nanopore tips yields stochastic emission bursts that map active binding sites with nanometric precision. By varying the fluorophore-metal distance using DNA spacers of controlled length, we observe a non-monotonic intensity response consistent with near-field quenching and plasmonic enhancement, identifying an optimal separation of around 6 nm. Finally, we extend the concept to dual-material Au/Si nanopores, demonstrating lateral coupling between plasmonic and semiconducting regions. These results establish DNA-PAINT as a quantitative probe of nanoscale optical environments in hybrid nanopores.

Keywords: plasmonic nanopores; DNA-PAINT; single-molecule fluorescence; near-field enhancement; fluorescence quenching; DNA nanorulers; dual-material nanopores; hybrid optoelectronic sensors; molecular biosensing

# Introduction

Solid-state nanopores have emerged as versatile platforms[1] for single-molecule sensors.[2,3] Their robustness, chemical stability, and compatibility with semiconductor processing have enabled several applications ranging from DNA sequencing[4,5] to iontronics[6,7] and even data-storage readout.[8,9] Following our earlier work on three-dimensional conical nanopores made of metal oxides, structures that established a robust and versatile platform for electrical and iontronic functionalities,[10] we now extend this concept toward plasmonic architectures that enable optical techniques at the single-molecule level. When metals such as gold are introduced, plasmonic antennas concentrate light into nanoscale hot spots,[11] enhancing excitation and emission rates[12] and enabling optical control of single-molecule dynamics,[13] gating,[14] and guiding chemical reactions at the nanoscale[15]. These plasmonic nanopores therefore provide a unique platform in which electromagnetic, electrical, and thermal fields can be tuned precisely within tens of nanometers.[16,17]

Understanding and exploiting this near-field regime requires a method capable of reporting local optical conditions with nanometer precision, ideally one that can reveal both spatial and temporal information on emission processes and kinetics. DNA-PAINT (Points Accumulation for Imaging in Nanoscale Topography) is an ideal candidate technique.[18] Based on the transient hybridization between short fluorescent "imager" strands and complementary "docking" sequences immobilized on a surface, DNA-PAINT produces stochastic fluorescence bursts that can be localized with sub-diffraction precision.[19,20] Its modular chemistry and single-molecule sensitivity make it particularly suitable for probing complex nanostructured interfaces,[21] including plasmonic architectures where localized fields and surface chemistry together determine optical behavior.[22]

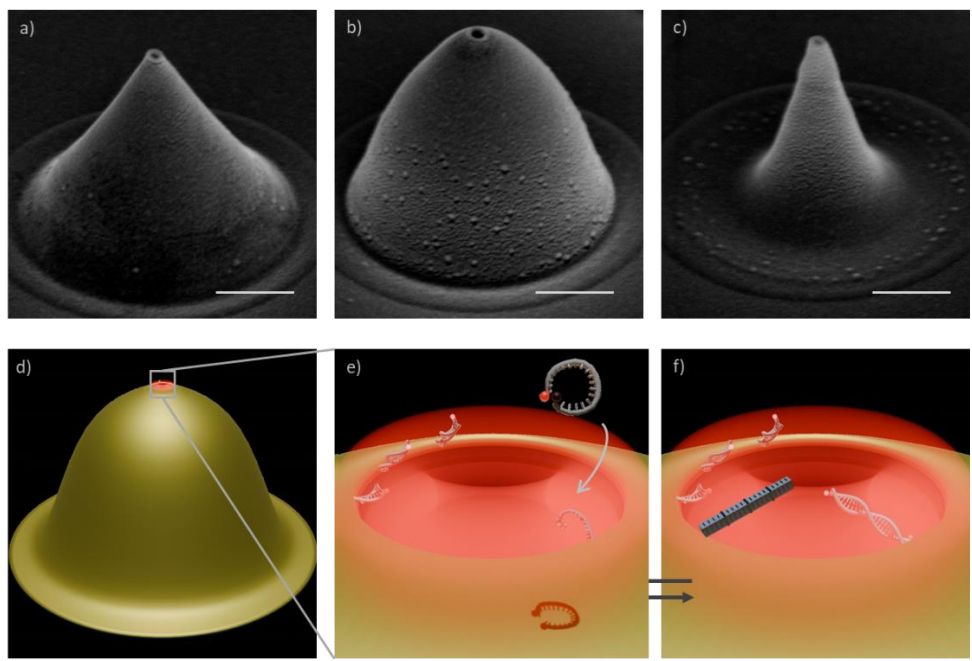

*Figure 1. Plasmonic nanopores and optical probing strategies. (top) SEM images of representative conical gold nanopores fabricated on thin membranes, showing reproducible tip morphology and nanoscale apertures at the center. Scale bars: 500 nm. (bottom) Schematic illustration of the two experimental configurations used in this work: DNA-PAINT imaging, where transient fluorescent events report local optical fields at the nanopore tips, and distance-dependent fluorescence measurements using DNA oligonucleotides of defined lengths (3 nm, 6 nm, 9 nm) acting as molecular nanorulers. Together, these approaches enable single-molecule mapping and quantitative analysis of plasmonic enhancement and quenching within the confined near-field region.*

In this work, we introduce DNA-PAINT and DNA-based nanorulers within plasmonic nanopores, establishing a new approach to probe optical fields and surface functionality at the single-molecule level. Fluorophores positioned at controlled distances from the metallic surface, through DNA oligos of different lengths, act as nanorulers to quantify how fluorophore-metal separation balances electromagnetic field enhancement and quenching. We find that the emission intensity follows a clear distance-dependent trend, with an optimal regime at intermediate spacing where enhancement dominates over non-radiative decay, in line with the typical maximum observed for emitters near gold nanospheres.[23] In parallel, transient DNA-PAINT binding events provide a dynamic optical readout localized at the nanopores, confirming selective functionalization and spatially confined fluorescence, indicating areas of higher brightness corresponding to plasmonic hotspots. Finally, we extend the concept to dual-material nanopores, and achieve hybrid plasmonic–semiconducting structures by laterally combining gold and silicon. The optical interaction between Au and Si at the nanoscale can itself enhance local fields,[24] and such asymmetric architectures could enable diverse surface chemistries and multimodal operation, opening a route toward multifunctional nanopore platforms. Figure 1 illustrates the architecture of the plasmonic nanopores and the combined experimental approaches used in this work. Conical gold nanopores fabricated on thin membranes present well-defined apertures at the tip, where DNA docking strands are immobilized for single-molecule DNA-PAINT readout. Fluorophores positioned at defined distances via short, medium, and long DNA oligos (3, 6, 9 nm) act as nanorulers to probe near-field effects. Together, these complementary strategies enable mapping of optical enhancement and quenching within the plasmonic nanopore geometry and establish the experimental framework for the results discussed below.

## Results and discussion

We first investigated how the fluorophore-metal separation distance influences emission strength in plasmonic nanopores by functionalizing identical pores with DNA oligonucleotides of three different lengths, therefore positioning the dyes at controlled separations from the metallic surface (Figure 2a).[25] Each sample was prepared by the same surface chemistry and labeling protocol to ensure that only the spacer length varied.

Using the three DNA spacers with different length, we observed a dependence of the fluorescence intensity on distance defined by the spacers in confocal microscopy as shown in Figure 2b-g (additional examples shown in Figure S1). The shortest linker (3 nm) produced weak emission, consistent with optical quenching by the metal; the intermediate 6 nm spacer maximized brightness, representing the regime where near-field enhancement dominates; and the longest 9 nm linker resulted in a slightly lower intensity, reflecting near-field decay with increasing separation.[26] Here, 3/6/9 nm denote nominal spacer lengths, and due to conformational tilting and flexibility the effective dye–metal separation is typically smaller than the nominal value.[27] Despite the three pore shapes producing distinct ionic behaviors,[10] their fluorescence readouts at 640 nm were indistinguishable, with no significant shape effect under our conditions.

Normalizing the fluorescence to the mean 6 nm intensity clarifies the relative efficiency of each spacer length: 3 nm yields ~0.15 ×⟨$I_{6nm}$⟩, 6 nm is set to 1, and 9 nm gives ~0.5 ×⟨$I_{6nm}$⟩. This non-monotonic response is consistent with the enhancement/quenching interplay near plasmonic antennas: 3 nm is expected to be strongly quenched, 6 nm maximizes enhancement, and 9 nm experiences weaker electromagnetic enhancement due to field decay in line with typical distance-dependent plasmon–emitter interactions.[23,28]

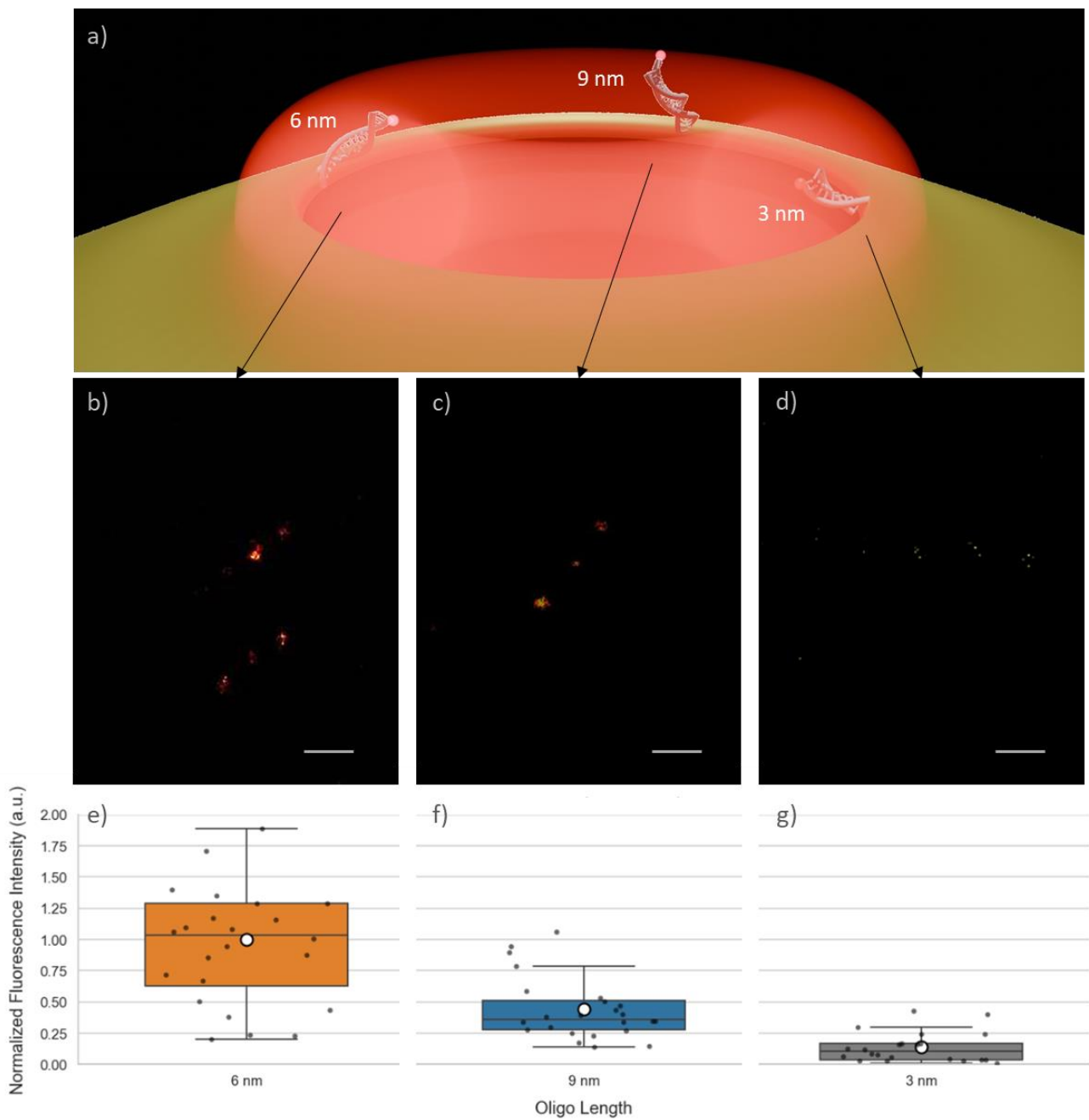

Figure 2. Fluorescence from plasmonic nanopores functionalized with DNA oligos of different lengths with Atto647N on the free end. (a) Schematic representation of a plasmonic nanopore showing fluorophores located at distinct distances from the metallic surface, corresponding to different near-field enhancement regions (red shading). (b-d) Confocal fluorescence maps of nanopores decorated with ~6 nm, 9 nm, and 3 nm oligos, respectively. Bright spots correspond to individual pores with distinct emission levels depending on the DNA oligo length. (e-g) Boxplots of the measured fluorescence intensities for each oligo length. Black dots represent individual nanopores and white circles indicate mean values. The non-monotonic intensity dependence confirms a balance between enhancement and quenching mechanisms, defining an optimal intermediate fluorophore-metal spacing for maximum brightness. Scale bars: 2 µm.

Variations in total intensity could also be influenced by differences in the number of fluorophores attached per nanopore. In addition to pure optical effects, mass transport and packing during deposition can bias total intensity. Shorter oligos have smaller hydrodynamic radii and higher diffusion coefficients, which facilitate reaching the surface over a fixed incubation time, and the adsorption/association rate scales with the diffusion coefficient, so faster diffusers deliver more strands to the pore within the same time.[29] Their smaller steric footprint can also permit denser packing in the confined rim region. Together, these factors increase the number of emitters N even if per-emitter brightness is reduced. A measurement capable of resolving the number of fluorophores per pore, such as single-molecule counting, or lifetime measurements,[30,31] would clarify whether the observed trends arise purely from optical distance effects or from variations in labeling density.

Overall, this systematic distance-dependent response provides design guidelines for plasmonic nanopore optimization. Avoiding the quenching zone directly adjacent to the metal, targeting intermediate fluorophore-metal separations where field enhancement dominates, and accounting for potential variations in labeling density are key for maximizing fluorescence yield. We therefore use simulations to map $|E|^2$ and identify where the enhancement maximizes.

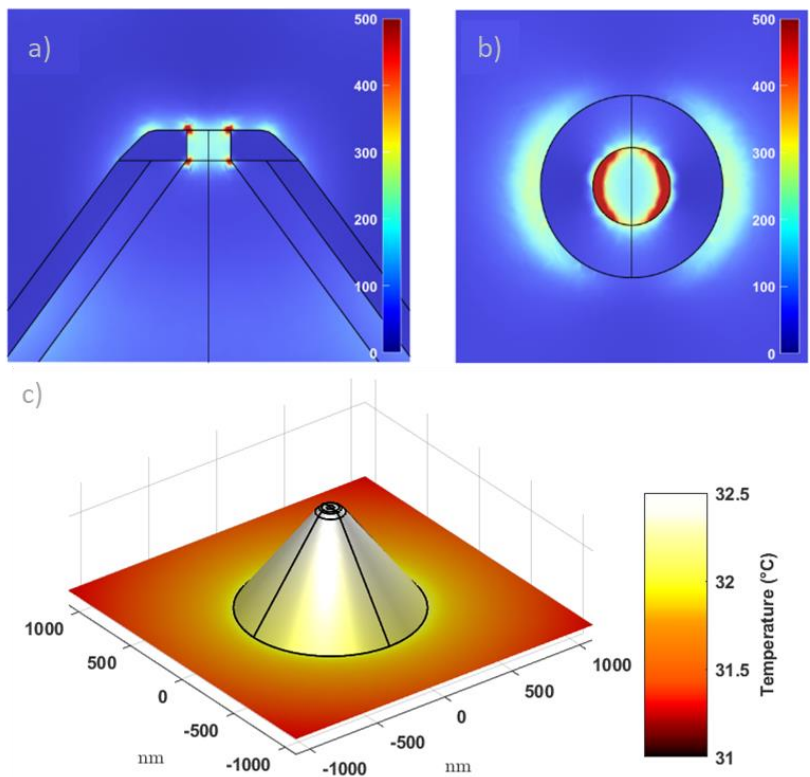

Figure 3. Simulated electric field enhancement ($|E|^2/|E_0|^2$) at λ = 634 nm for 40 nm 3D nanopores. (a, b) Cross-sectional and top-view maps of the field intensity for a gold-coated conical nanopore. The excitation wavelength (634 nm) corresponds to the peak in the simulated plasmonic enhancement, close to the

*experimental illumination at 640 nm. The mode produces strong, symmetric confinement around the pore, with maximum intensity within a few nanometers from the metal surface. (c). Steady-state temperature map at λ = 634 nm, irradiance = 15 µW µm⁻² (≈1.5×10⁷ W m⁻²), bulk $T_o$ = 20 °C; color scale shows T.*

As shown in Figure 3 a-b) the field is strongly enhanced within the first few nanometers from the metal surface, and then the field rapidly decays after, and emitters positioned 5–7 nm from the metal sample the maximum $|E|^2$ region, while avoiding quenching. [32] In contrast to the electrical characteristics previously reported for similar 3D conical nanopores,[10] our simulations did not reveal a significant dependence of the near-field distribution on the wall angle of the conical structure.

Plasmonic metals irradiated at resonant wavelengths do not only lead to a strong increase in the local electromagnetic field, but also produce significant thermal losses. This is because a certain amount of the absorbed energy is converted into heat, and therefore a noticeable increase in temperature near the metal nanostructures occurs (Figure 3 c).

In the nanopore geometry the thermal contribution is particularly significant, as the area coated by the plasmonic metal (gold) is very large, and fully encompasses the hot spot region at the pore aperture. Figure 3c shows that our calculations yield a temperature of approximately 32.5°C around the nanopore under irradiation with a laser at a wavelength of 635 nm and 15 µW/µm², from a bulk temperature of 20°C. Such local heating can become crucial for experiments involving biomolecules and DNA.

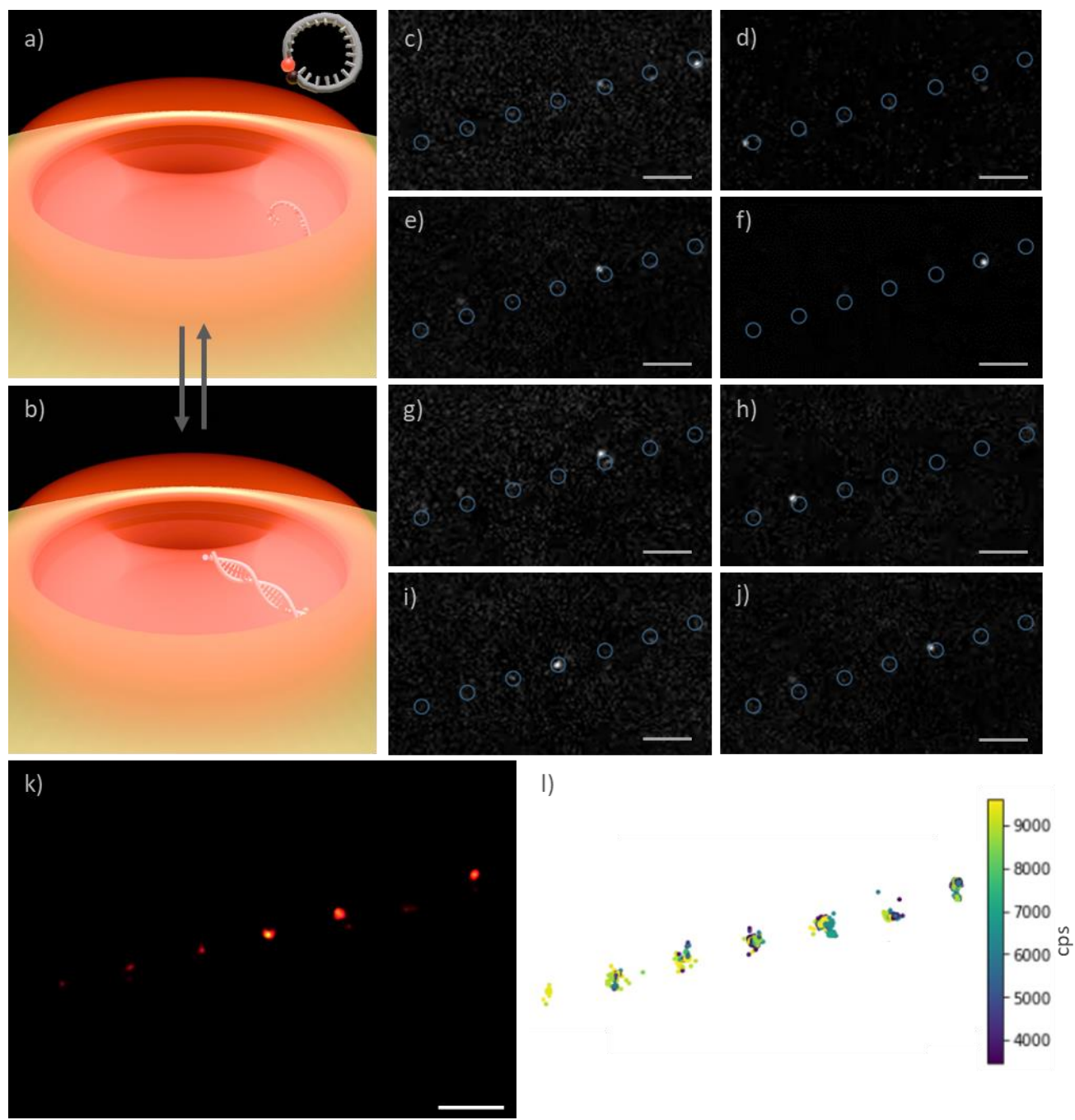

*Figure 4. DNA-PAINT on plasmonic nanopores. (a,b) Schematic representation of the DNA-PAINT process occurring inside a plasmonic nanopore. Transient hybridization between fluorescently labeled imager strands and complementary docking sequences at the pore tip generates stochastic fluorescence bursts. (c-j) Snapshots from wide-field fluorescence imaging, showing individual binding events occurring on an array of nanopores (highlighted by blue circles). (k) Localization map reconstructed with Picasso,[19] displaying the positions of single-molecule events accumulated over several minutes. (l) Emission rate (counts per second) per localization, revealing the distribution of emission intensities across nanopores. The enhanced brightness in selected pores indicates local electromagnetic field amplification in or near the nanopore. Scale bars: 2 μm.*

DNA-PAINT imaging was then used to probe the optical response of plasmonic nanopores functionalized with DNA docking strands. Upon introducing complementary fluorescent imager strands in solution, transient hybridization events produce characteristic fluorescence bursts at the nanopore openings (Figure 45 a,b). Unlike conventional DNA-PAINT experiments performed at the glass–water interface in total internal reflection fluorescence (TIRF), here the active nanostructures are located several hundred nanometers above the coverslip, within the metallic membrane, as shown in Figure S2. Consequently, evanescent illumination cannot selectively excite bound imagers (as in TIRF), and measurements were carried out under wide-field epi-illumination conditions. To suppress fluorescence background from freely diffusing strands, we employed fluorogenic imager oligonucleotides carrying a dye–quencher pair that remain dark in solution, and fluoresce only upon hybridization to the docking sequence at the nanopore tip.[32] The temporal nature of these events enables single-molecule localization and super-resolution reconstruction. At our chosen oligo length (15 bp imager) and attachment geometry, the fluorophore is expected to stand ~5-7 nm from the metal surface,[27] within the region of maximum near-field enhancement identified earlier.

Sequential frames (Figure 4 c-j) show discrete, diffraction-limited spots corresponding to single binding events occurring on individual nanopores. The consistent alignment of these events with the predefined array confirms that hybridization is spatially confined to the tips of the nanopores. Each pore acts as a nanometric reaction site, where the local optical field modulates the brightness of single binding events, and the event frequency is defined by the DNA-PAINT kinetics set by the imager concentration and sequence. Minor variations between pores arise from differences in local accessibility, diffusion, or temperature variations near the tips.

Localization analysis using Picasso[19] (Figure 4k) revealed dense clusters of events at the nanopore positions. The reconstructed map confirms that DNA-PAINT signals originate exclusively from the pores, with minimal background elsewhere on the substrate. Notably, despite the overnight incubation applied to the entire surface, we did not observe meaningful PAINT signal on the surrounding flat substrate when focusing the optics on the plane of the substrate (Figure S3). This confirms that our protocol[12] is suitable to investigate the localized effects at the nanopore apex.

To quantify the emission characteristics, we first examined the integrated photon count per event, i.e. the total number of photons detected during each transient hybridization (Supporting Information, Figure S4). The integrated signal depends on both brightness and event duration. The imager-docking pair used here was designed to yield characteristic binding durations of about one second, consistent with standard DNA-PAINT kinetics.[32,33] In our data, dwell times ranged from ~0.1 s to >10 s, corresponding to total photon yields between $10^3$ and $5 \times 10^5$ photons per event. The correlation between photon count and dwell time confirms that long-lived events, even if moderately bright, can accumulate photon totals even larger than short, intense bursts (Figure S4).[34]

From these quantities, we derived the emission rate (counts per second) for each event to visualize spatial variations across the nanopore array (Figure 5l). Each cluster corresponds to a single nanopore and displays characteristic intensity levels, typically between $4 \times 10^3$ and $9 \times 10^3$ photons s$^{-1}$ (5-95 % range). The resulting map highlights heterogeneity in emission rate and brightness, suggesting local differences in excitation field strength, molecular orientation, or nanoscale geometry at the pore tips. The fitted background showed no correlation with event intensity (r < 0.1; see Figure S5), confirming that brightness variations are not due to local background but rather originate from the plasmonic structures and localized field enhancement. [19,35]

Conversely, enhanced local heating in high-field regions could accelerate unbinding, leading to shorter dwell times but higher brightness. [36] However, dwell times represent true binding kinetics only if the dye remains emissive throughout the bound state. Under strong excitation, photobleaching can prematurely terminate fluorescence, particularly near plasmonic hot spots where local field is strongest. As a result, apparent unbinding events may partly reflect bleaching rather than dissociation. Distinguishing these effects will require extended analyses of possible thermoplasmonic influences [37,38] as described for similar plasmonic nanopore systems,[12] . The corresponding dwell-time distributions are reported in Figure S4, but further studies with controlled excitation and larger datasets will be necessary to disentangle these factors.

These results demonstrate that DNA-PAINT can serve not only as a molecular imaging technique but also as a sensitive reporter of the local nanoenvironment within plasmonic structures.[39] Importantly, this represents the first experimental realization of DNA-PAINT within solid-state nanopores, establishing a new platform that unites single-molecule localization microscopy with nanofluidic confinement. Future experiments combining photon-count statistics, dwell-time distributions, and temperature mapping will help to clarify the relative contributions of optical enhancement, local heating, and kinetic effects in shaping the observed fluorescence dynamics.

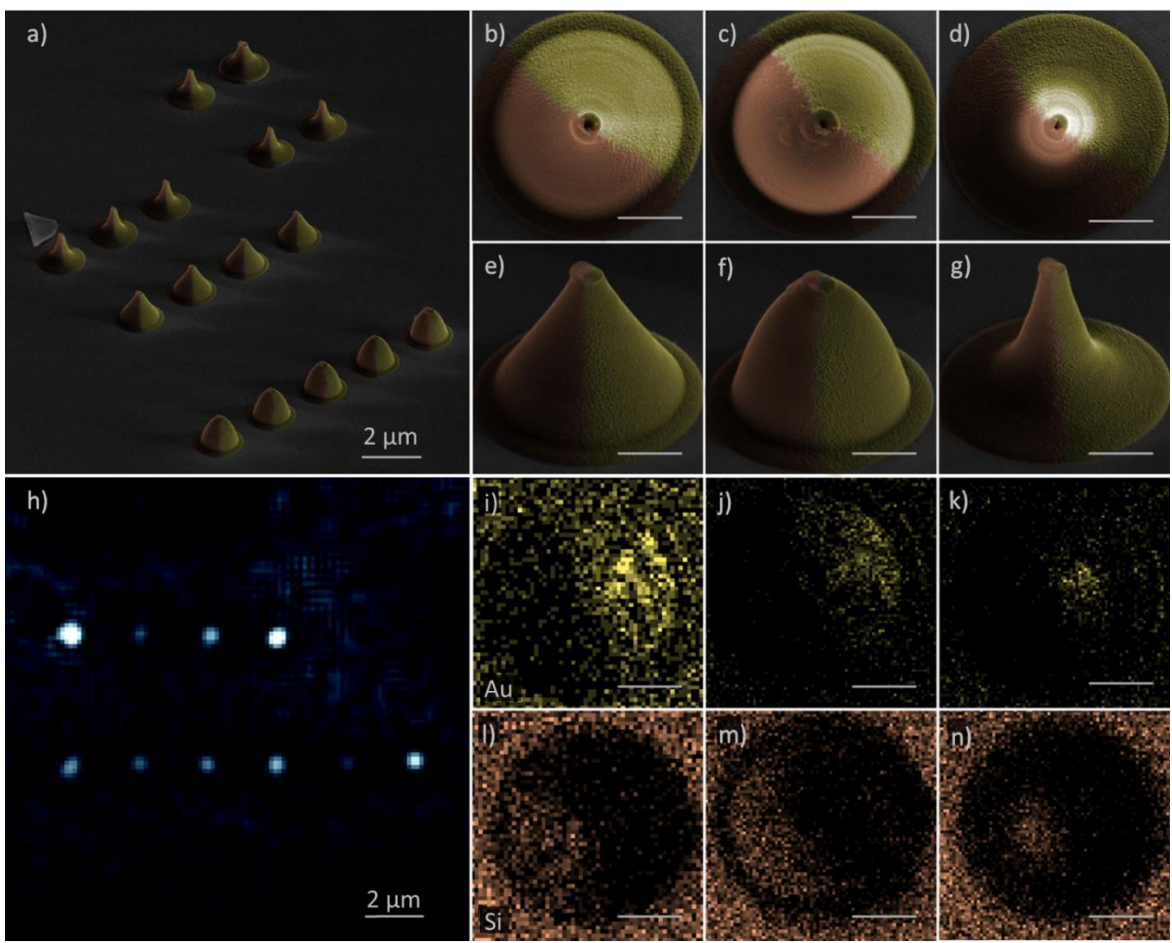

*Figure 5. Dual-material nanopores: morphology, fluorescence, and composition. (a) SEM image of an array of dual-material nanopores with varying geometries and (b-g) high-magnification SEM views of individual pores showing distinct cone profiles obtained by adjusting the fabrication parameters, false-colored to highlight the gold (yellow) and silicon (orange) regions, illustrating the lateral Au/Si junction. (h) Confocal fluorescence map of the same nanopore array functionalized with 6 nm Atto647-DNA oligos, showing localized emission at the pore tips. (i-n) EDX elemental mapping of the dual nanopores: gold (yellow) and silicon (orange) channels highlight the lateral composition of the hybrid structure, confirming the spatial segregation of the plasmonic and semiconducting regions. Scale bars: 2 μm in (a, h); 500 nm in (b-g, i-n).*

Finally, we explored how introducing a second material could expand the design space for hybrid optical and electronic responses.[40] For this purpose, we fabricated dual-material conical nanopores in which gold and silicon halves are laterally joined within the same structure, allowing a plasmonic material and a high-index semiconductor to interact (Figure 5).[41,42]

The SEM images demonstrate well-defined Au/Si interfaces and reproducible pore geometries ranging from shallow cones to sharper tips. Upon functionalization with short (6 nm) DNA spacers carrying fluorophores, confocal mapping reveals localized emission at the pore tips, consistent with our decoration method and coupling to the plasmonic resonance.[43] EDS elemental maps further confirm the lateral segregation between gold and silicon, validating the hybrid design. In the dual nanopore (Figure S7), the gold-silicon interface introduces a pronounced asymmetry: the near field remains concentrated on the metallic side while the silicon region shows negligible enhancement at this wavelength.[44] This configuration enables directional control of excitation and potentially multi-modal operation by balancing the optical responses of the two materials.

These dual-material nanopores introduce a versatile route toward multifunctional single-molecule devices, combining optical field enhancement with the electrical and surface tunability of semiconductors. Their asymmetric composition could also enable directional or chiral optical responses,[45] opening opportunities for advanced photonic and sensing applications.

## Conclusions

We have demonstrated, for the first time, the implementation of DNA-PAINT within plasmonic nanopores, introducing a single-molecule optical readout directly coupled to a nanofluidic system. The transient binding of fluorescent imagers at the nanopore tips confirmed selective tip functionalization and enabled mapping of localized field enhancement. This work complements the previously explored electronic functionalities of three-dimensional conical nanopores by adding an optical pathway to probe and quantify electromagnetic field enhancement at the nanoscale.

By employing DNA spacers of controlled length, we further quantified how the fluorophore-metal separation regulates the balance between electromagnetic enhancement and quenching. The emission trend (minimal brightness at 3 nm, a maximum at 6 nm, and reduced intensity at 9 nm) defines a practical design window for optimizing optical response in metallic nanopores. These results highlight that emitters placed within an intermediate distance regime experience maximal enhancement, while those closer to the surface are dominated by non-radiative losses.

Finally, the introduction of dual-material Au/Si nanopores expands the concept toward multifunctional architectures. By combining plasmonic and semiconducting regions within a single conical geometry, these structures offer simultaneous optical and electronic functionality, enabling tailored field distributions, selective chemistries, and potentially chiral optical responses.

Collectively, our results establish plasmonic nanopores as a powerful nanoscale platform for correlating optical fields, thermal effects, and molecular kinetics in confined geometries. The integration of DNA-based probes provides a strategy to engineer and explore nanostructured interfaces with precision, opening new directions for single-molecule sensing, plasmon-enhanced spectroscopy, and DNA data-storage technologies. Future work will correlate fluorescence lifetime, local temperature, and ionic conductance to fully map coupled optical-electrical effects in hybrid nanopores.

# Experimental Methods

Substrate and Nanopore Fabrication.

Three-dimensional oxide nanopores were fabricated on SiN/Si substrates by combining spin coating of photoresist, focused ion beam (FIB) drilling, atomic layer deposition (ALD) of dielectric material ($SiO_2$ or $Al_2O_3$), and a final development step[10]. A 20 nm Au layer was then deposited to form a plasmonic nanoantenna at the pore tip, with a 3-5 nm Cr or Ti adhesion layer between the dielectric and metal. Dual-material nanopores were obtained by directional metal deposition: Au was evaporated at 90° to coat one side of the conical pores, followed by Si deposition after rotating the substrate 180°, yielding laterally asymmetric dielectric-plasmonic structures.

Surface Chemistry and Tip-Selective Functionalization.

Functionalization followed an established nanopore decoration protocol[12] adapted to DNA oligonucleotides in triethanolamine (TEA) buffer. Chips were plasma-activated for 5 min from the back (dielectric) side, TEA buffer was added to the top side, and a DNA-linker solution was applied on the back side. Docking strands or oligo nanorulers were incubated either overnight (DNA-PAINT) or briefly (0.5-2 min) for distance-series labeling. Samples were rinsed in TEA buffer. Under these asymmetric conditions, oligonucleotide attachment localized predominantly at the nanopore tip with negligible adsorption on the surrounding flat substrate.

DNA-PAINT.

Docking strands (Table S1) were incubated overnight unless stated otherwise. Imaging strands were introduced during acquisition under standard PAINT conditions at concentrations between 50 pM and 50 µM. Tip and base regions were imaged separately by shifting the focal plane 1-2 µm. Short-incubation tests (minutes) were performed for comparison with nanoruler experiments.

Imager: 5'-/Atto633N/AAGTTGTAATGAAGA/BHQ$_2$/-3'

Docking: 5'-/ThiolMC6-D/TTATCTCCTATACAACTTCC/-3'

Oligo Nanorulers.

Three double-stranded DNA constructs positioned the fluorophore at around 3, 6, and 9 nm from the metal through defined poly-A/T spacers. After short incubations (30 s-1 min), confocal fluorescence maps were recorded around individual pores.

3 nm: 5'-/ThiolMC6-D/AAAAAAAAAA/Atto647N/-3' + 5'-/TTTTTTTTTT/-3'

6 nm: 5'-/ThiolMC6-D/AAAAAAAAAAAAAAAA/Atto647N/-3' + 5'-/TTTTTTTTTTTTTTT/-3'

9 nm: 5'-/ThiolMC6-D/AAAAAAAAAAAAAAAAAAAAA/Atto647N/-3' + 5'-/TTTTTTTTTTTTTTTTTTTT/-3'

Optical Measurements.

Fluorescence intensity maps of nanoruler-functionalized nanopores were acquired on a Nikon A1R+ confocal microscope using a 100× oil-immersion objective (NA 1.49) and 640 nm excitation (50 µW). Emission was collected through standard Cy5 filter sets and recorded under identical conditions for all samples to enable quantitative comparison of fluorescence enhancement versus distance.

DNA-PAINT measurements were performed on a custom-built wide-field fluorescence microscope based on an inverted Olympus IX83 body. Samples were illuminated in epi-configuration with a 640 nm laser (Laser Quantum) providing a Gaussian beam and circularly polarized excitation. Fluorescence was collected through a 100× oil-immersion objective (NA 1.5, Olympus), filtered by a 532/640 nm dual-band set (Chroma), and detected on a CMOS camera (ORCA-Fusion, Hamamatsu).

Numerical Simulations.

Three-dimensional electromagnetic simulations were performed in COMSOL Multiphysics using realistic conical geometries including the dielectric and Au layers. Near-field intensity distributions were computed under 640 nm excitation to identify zones of field enhancement.

ACKNOWLEDGMENTS

The authors acknowledge funding from the European Union under the Horizon 2020 Program, FET-Open: DNA-FAIRYLIGHTS, Grant Agreement 964995, the European Union Program HORIZON-Pathfinder-Open: 3D-BRICKS, grant Agreement 101099125, the HORIZON-MSCADN-2022: DYNAMO, grant Agreement 101072818. The authors thank the Clean Room Facility of IIT for the support in sample fabrication.